\documentstyle[12pt]{article}
\begin{document}
\setlength{\baselineskip}{0.30in}
\newcommand{\beq}{\begin{equation}}
\newcommand{\eeq}{\end{equation}}
\newcommand{\bi}{\bibitem}

{\hbox to\hsize{March, 1997  \hfill TAC-1997-009}
\begin{center}
\vglue .06in
{\Large \bf {Instructive Properties of
Quantized Gravitating Dust Shell
  }
}
\bigskip
\\{\bf A.D. Dolgov}
 \\[.05in]
{\it{Teoretisk Astrofysik Center\\
 Juliane Maries Vej 30, DK-2100, Copenhagen, Denmark
\footnote{Also: ITEP, Bol. Cheremushkinskaya 25, Moscow 113259, Russia.}
}}\\
{\bf I.B. Khriplovich }
 \\[.05in]
{\it{ Budker Institute of Nuclear Physics\\
  630090 Novosibirsk, Russia
 }}\\[.40in]

\end{center}
\begin{abstract}
We investigate quantum dynamics of self-gravitating spherical dust shell.
The wave functions of discrete spectrum are not localized inside the
Schwarzschild radius. We argue that such shells can transform into white
holes (in another space). It is plausible that shells with bare masses
larger than the Planck mass loose their mass emitting lighter shells.
\end{abstract}

Thin dust shell is one of the simplest models of collapsing gravitating bodies.
Equations of motion of such objects were derived in ref. \cite{isr1}. The
classical dynamics of this system was considered in
refs. \cite{kuch,bkt,bg}. This model was quantized in various
nonequivalent ways with physically different results
in papers \cite{bkkt,ber1,hkk,ber2}. In our opinion, the most natural
approach was proposed in ref. \cite{hkk} where the problem was reduced to the
usual $s$-wave Klein-Gordon equation in a Coulomb field. In this note
we would like to point out some curious effects arising in this approach
which may turn instructive for more realistic situations.

Let us briefly present the basic steps leading to the Klein-Gordon equation
for this system. The situation is rather unusual because there are two
classical equations of motion for the shell radius $r$ \cite{isr1}:
\beq{
{\ddot r + km /r^2 \over \sqrt{ 1+\dot r^2 -2km/r}} +
 {\ddot r\over \sqrt{ 1+\dot r^2} } = 0;
\label{ddotr1}
}\eeq
\beq{
{\ddot r + km /r^2 \over \sqrt{ 1+\dot r^2 -2km/r}} -
 {\ddot r\over \sqrt{ 1+\dot r^2} } = {k\mu \over r^2}.
\label{ddotr2}
}\eeq
These two equations for a single variable are consistent, as will be
demonstrated below, since they depend on the total mass $m$ (or the
total energy) of the shell as defined by the
Schwarzschild metric outside the shell. It means in fact that the
equations depend on the initial conditions. In these equations
$\mu $ is the "bare" mass of the shell,
i.e. the rest mass of each
particle of the dust times their number, $k$ is the Newton gravitational
constant, and $\dot r = dr/d\tau$ where $\tau$ is the proper time of the dust.
First integrals of these equations can be written respectively as
\beq{
\sqrt{ 1+\dot r^2 -2km/r}\, +\, \sqrt{  1+\dot r^2} \,= C_1;
\label{int1}
}\eeq
\beq{
\sqrt{ 1+\dot r^2 -2km/r}\, -\, \sqrt{  1+\dot r^2} \,= C_2 -{k\mu\over r}.
\label{int2}
}\eeq
Multiplying these two expressions we obtain:
\beq{
{2km \over r } = C_1 \left(  { k\mu\over r} - C_2 \right)
\label{int3}
}\eeq
>From the last equality it follows that the integrals (\ref{int1}) and
(\ref{int1}) are compatible only if $C_2 =0$ and $C_1 = 2m /\mu$.
If these conditions are imposed, the equations of motion (\ref{ddotr1}) and
(\ref{ddotr2}) are consistent and equivalent to
\beq{
{\ddot r\over \sqrt{ 1+\dot r^2} } = - {k\mu \over 2r^2}.
\label{ddotr3}
}\eeq
Equation (\ref{ddotr3}) has the following first integral
\beq{
m = \mu \sqrt{  1+\dot r^2}  - {k\mu^2\over 2r}.
\label{ham1}
}\eeq
This expression will be taken as a classical Hamiltonian of the system
(after going over from the velocity $\dot r$ to the canonical momentum $p$).
>From the formal point of view
(of the theory of differential equations, or analytical mechanics) any
function of a first integral can be chosen as a Hamiltonian.
The choice is by no means unique,
but eq. (\ref{ham1}) is singled out because this Hamiltonian corresponds to
the total energy of the shell, and this energy $m$ has an explicit and
simple expression.

One can easily recognize in the rhs of eq. (\ref{ham1})
the energy of a relativistic particle in a Coulomb-like field,
$-k\mu^2/ 2r$, written in proper time $\tau$. It is convenient to go
over from $\tau$ to the world time $t$ inside the shell:
$$
d\tau = dt \sqrt{ 1 - v^2 },\; v = dr/dt.
$$
We obtain now:
\beq{
m = {\mu \over \sqrt{  1- v^2}}  - {k\mu^2\over 2r}.
\label{ham2}
}\eeq
Clearly in this case the canonical momentum equals $ p= \mu v /\sqrt{1-v^2}$
and the Hamiltonian has the well known form:
\beq{
H =  \sqrt{ p^2 + \mu^2}  - {k\mu^2\over 2r}.
\label{ham3}
}\eeq

The quantum-mechanical wave equation corresponding to this Hamiltonian is
derived by the standard procedure, taking the square of the root,
i.e., rewriting (\ref{ham3}) as:
 $$
(H + k\mu^2/ 2r)^2 = p^2 + \mu^2.
$$
Thus one obtains the usual Klein-Gordon radial
equation for $s$-wave:
\beq{
\left( \partial_r^2 + {2\over r}\, \partial_r + {k\mu^2 m \over r}
+ {k\mu^2 \over 4 r^2} + m^2 -\mu^2  \right) \psi = 0.
\label{kg}
}\eeq

The discrete spectrum of this equation is well-known \cite{som,bet}:
\beq{
m_n = \mu \left[ 1+ { k^2 \mu^4 \over
\left( 2n + 1  + \sqrt{1 - k^2 \mu^4 } \right)^2 } \right]^{-1/2}
\label{mn}
}\eeq
The radial quantum number $n$ is integer and runs from 0 to infinity.

The spectrum has a singularity at $k\mu^2 =1$. At larger values of $\mu$ the
$r^{-2}$ potential becomes so strong that the "fall to the center" takes place,
i.e. there are no stationary states. It looks natural that for heavy
pressureless matter (with $\mu > m_{Pl} = 1/\sqrt{ k}$) naive quantum
mechanical effects cannot stop the collapse.

The curious property of the states belonging to the discrete spectrum should be
pointed out. Even in the most tightly bound ground state for $k \mu^2 =1$ the
wave function is not localized inside the gravitational radius of the shell.
The probability to find the shell outside it
is $3/e^2 \approx 0.4$. Here we naively
use $r$ as the operator of coordinate. In the relativistic case a more
refined definition of the coordinate should be used \cite{sch,hkk}. It is
clear however that any reasonable definition of the
coordinate operator would not change the localization considerably.

Let us consider now the continuous spectrum. Though we assume as above
that  $k\mu^2 <1 $ or in other words $\mu < m_{Pl}$, the total energy $m$
can be arbitrarily large.
The wave function of such a state
is a superposition of incoming and outgoing spherical waves with equal
amplitudes. It is evidently non-localized. To consider the quantum analogue
of the collapse of the classical shell we have to turn to wave packets.
As was noted in ref. \cite{hkk} the gravitational radius of this object will
be as smeared as its energy. Let us assume that
the initial radius of the maximum of the incoming spherical
wave packet is much larger than its average gravitational radius $r_g$.
>From the point of view of a distant observer this packet moving towards
the center freezes at $r=r_g$. However, in its proper time  it
reaches the center in a finite interval $\delta\tau$, bounces back and then
after the same time interval returns to its initial position and form.
Certainly, it
returns not to "our" space, but to a quite different one.
This is a possible realization of the white hole phenomenon \cite{idn,yn}.

The considered realization of a white hole based on quantum scattering differs
essentially from the classical examples of white holes. In the classical case
the very existence of the phenomenon depends crucially upon the presence of
singularity at $r=0$, while in our case the transformation of an incoming
spherical wave into an outgoing one takes place even for nonsingular
potentials.

Finally let us return to the case of a large bare mass, $\mu >m_{Pl}$.
This problem formally coincides with that of a charged scalar particle
in the field of a point-like nucleus with a supercritical electric charge,
$Z\alpha > 1$. It is known that the vacuum around a supercharged nucleus is
unstable and this nucleus discharges by emitting positively charged
particles (see e.g. refs. \cite{pz,mig}).
A
similar scenario is plausible for
our problem: the collapsing shell loses its bare
mass $\mu$ by emitting light shell-lets till it reaches the subcritical
mass. This phenomenon would resemble
quantum evaporation of usual black holes. On the other hand, in the
subcritical situation, $\mu < m_{Pl}$, the emission of shell-lets
does not take place (even if the physical mass $m$ is larger than
$m_{Pl}$). It can be considered as a hint that small black
holes do not evaporate, though at $m > \mu$ they
form white holes in another universe.

\bigskip
{\bf Acknowledgments}
We are grateful to I.D. Novikov for helpful discussions.
I.Kh. thanks TAC for hospitality.
The work of A.D. was supported in part by the Danish National Science
Research Council
through grant 11-9640-1 and in part by Danmarks Grundforskningsfond through its
support of the Theoretical Astrophysical Center. I.Kh. acknowledges the
support by the Russian Foundation for Basic Research through grant
No. 95-02-04436-a.

\end{document}